\documentclass[twocolumn]{svjour3}

\usepackage{graphicx}
\usepackage[square,sort,comma,numbers]{natbib}

\begin{document}

\title{Modulation of superconducting transition temperature in LaAlO$_{3}$/SrTiO$_{3}$ by SrTiO$_{3}$ structural domains}

\author{Hilary Noad \and Pascal Wittlich \and Jochen Mannhart \and Kathryn A. Moler}

\institute{
Hilary Noad \and Kathryn A. Moler 
\at Stanford Institute for Materials and Energy Sciences, SLAC National Accelerator Laboratory, 2575 Sand Hill Road, Menlo Park, CA 94025, USA\\
\email{Hilary.Noad@cpfs.mpg.de}
\and
Pascal Wittlich \and Jochen Mannhart 
\at Max Planck Institute for Solid State Research, 70569 Stuttgart, Germany
}

\maketitle

\begin{abstract}
The tetragonal domain structure in SrTiO$_{3}$ (STO) is known to modulate the normal-state carrier density in LaAlO$_{3}$/SrTiO$_{3}$ (LAO/STO) heterostructures, among other electronic properties, but the effect of STO domains on the superconductivity in LAO/STO has not been fully explored. Using a scanning SQUID susceptometer microscope to map the superconducting response as a function of temperature in LAO/STO, we find that the superconducting transition temperature is spatially inhomogeneous and modulated in a pattern that is characteristic of structural domains in the STO.

\end{abstract}

\section{\label{sec:introduction}Introduction}
Since the discovery of superconductivity at the \\{LaAlO$_{3}$/SrTiO$_{3}$} (LAO/STO) interface \cite{ReyrenSci07}, a variety of effects related to the superconductivity have been reported, including a dome in the superconducting transition temperature, $T_{c}$, as a function of backgate voltage \cite{CavigliaNature08}, coexistent superconductivity and magnetism \cite{DikinPRL11, LiNatPhys11, BertNatPhys11}, and pseudogap-like behavior in tunneling spectra \cite{RichterNat13}. LAO/STO has held great promise as a platform for studying low-dimensional superconductivity, as the gate tunability of the carrier density and $T_{c}$ \cite{CavigliaNature08} eliminates the need for chemical doping, raising the possibility of obtaining a clean (i.e. low disorder), two-dimensional system. Such a system is of interest for studying intrinsic effects such as the Berezinskii-Kosterlitz-Thouless phase transition \cite{BerezinskiiSovPhysJETP71, BerezinskiiSovPhysJETP72, KosterlitzJPhysC73}.

In tandem with the work on LAO/STO mentioned above, evidence has emerged from studies of superconductivity in 2D-doped STO \cite{NoadPRB16}; superconducting films of Nb, NbN, and underdoped YBa$_{2}$Cu$_{3}$O$_{7-{\delta}}$ grown on STO \cite{WissbergPRB17}; and the normal-state conductivity and electrostatic potential of LAO/STO \cite{FrenkelNatMater17, KaliskyNatMater13, HonigNatMater13} that tetragonal domain structure in the STO causes spatial variation in the electron system under study in each of those materials. Supplemental data in \cite{KaliskyNatMater13} suggested that STO domains modulated the superconducting response of the LAO/STO system below $T_{c}$, but the temperature dependence of the modulation and any possible influence on $T_{c}$ remained unexplored.

Here we use dilution fridge scanning superconducting quantum interference device (SQUID) susceptometry \cite{BjornssonRevSciInstrum01, HuberRevSciInstrum08} to directly image the micron-scale spatial variation of superconductivity in LAO/STO. We show that the superfluid density and transition temperature of the superconductivity at the interface are also modulated by tetragonal domain structure of the STO. 

\section{\label{sec:experimental}Experimental Setup}
We measured two LAO/STO heterostructures grown under nominally identical conditions on TiO{$_{2}$}-terminated {(001)}-oriented {SrTiO$_{3}$} substrates using pulsed laser deposition. The samples differ in the oxygen partial pressures used during post-growth annealing, as summarized in  in Table \ref{table:conditions}. Sample A was annealed in conditions chosen to minimize the contribution of oxygen vacancies to the conductivity, such that the two-dimensional electron system at the LAO/STO interface and its properties would be dominated by the polar catastrophe mechanism. Sample B was annealed in highly reducing conditions in order to introduce oxygen vacancies.

\begin{table}
\centering
\caption{\label{table:conditions}Post-growth annealing conditions for the two LAO/STO samples studied here. Both samples consisted of 6 unit cells of LAO deposited by pulsed laser deposition on {TiO$_{2}$}-terminated {(001)}-oriented {SrTiO$_{3}$} substrates. They were grown at 640$^{\circ}$C in $8{\times}10^{-4}$ mbar O$_{2}$. Each annealing step took 40 minutes.}
\begin{tabular}{l p{1.5cm} p{1.5cm} p{1.5cm}}
\\
\hline
Sample  & \multicolumn{3}{c}{mbar O$_{2}$}\\
  & 600$^{\circ}$C & 500$^{\circ}$C & 400$^{\circ}$C\\
\hline
A & 4  & 40 & 400 \\
B & $1{\times}10^{-5}$ & $1{\times}10^{-5}$  & $1{\times}10^{-5}$\\
\hline
\end{tabular}
\end{table}

 We measured the diamagnetic response of the superconductivity using a scanning SQUID susceptometer \cite{HuberRevSciInstrum08} by applying an ac magnetic field with the susceptometer's field coil and performing a lock-in measurement on the flux produced in response to the applied field. 

All of the susceptibility data presented here were taken using a field coil current of 0.25 mA$_{rms}$, resulting in an ac field of approximately $0.25$ G at the sample surface.  The background subtraction was determined by measuring the susceptibility away from the sample for each imaged area.

We mounted the samples with their $\langle$100$\rangle$-oriented edges approximately aligned with the X and Y axes of our scanner. We affixed the insulating STO substrate to copper tape with conducting silver paint (GC Electronics - Silver Print II, Part No. 22-023) and made aluminum wirebonds to the conducting interface with a standard ultrasonic wedge wirebonder (Westbond, model 7476E).  An electrically insulating, thermally conducting sapphire plate separated the copper tape of the backgate from the copper sample mount plate underneath. The sample mount was firmly attached to another copper piece in the microscope cage, which hangs from the mixing chamber, for good thermal contact. The backgate and sample contact remained grounded for all measurements presented here.

We measured three spatially separated regions on both Sample A and Sample B, totalling approximately $5.83{\times}10^{4}$ ${\mu}$m$^{2}$ and $4.35{\times}10^{4}$ ${\mu}$m$^{2}$, respectively, or about 1\% of the total $2{\times}2$ mm$^{2}$ surface area in both cases.  For both samples, two of the areas measured $76 \times 68$ ${\mu}$m$^{2}$. The third area on Sample A measured approximately $214 \times 224$ ${\mu}$m$^{2}$ and the third area on Sample B measured approximately $213 \times 156$ ${\mu}$m$^{2}$. On Sample A, successive areas were separated by 152 ${\mu}$m and 456 ${\mu}$m, respectively, along the direction of shortest separation; on Sample B, successive areas were separated by 706 ${\mu}$m and 412 ${\mu}$m, respectively, along the direction of shortest separation. 

\section{\label{sec:results}Results}

Sample A and B both display inhomogeneous superconductivity at tens of mK, but the two samples differ in the spatial distribution of their superconducting patches and in the spatial motifs of regions that exhibit locally lower $T_{c}$. 

\begin{figure}
\centering
\includegraphics{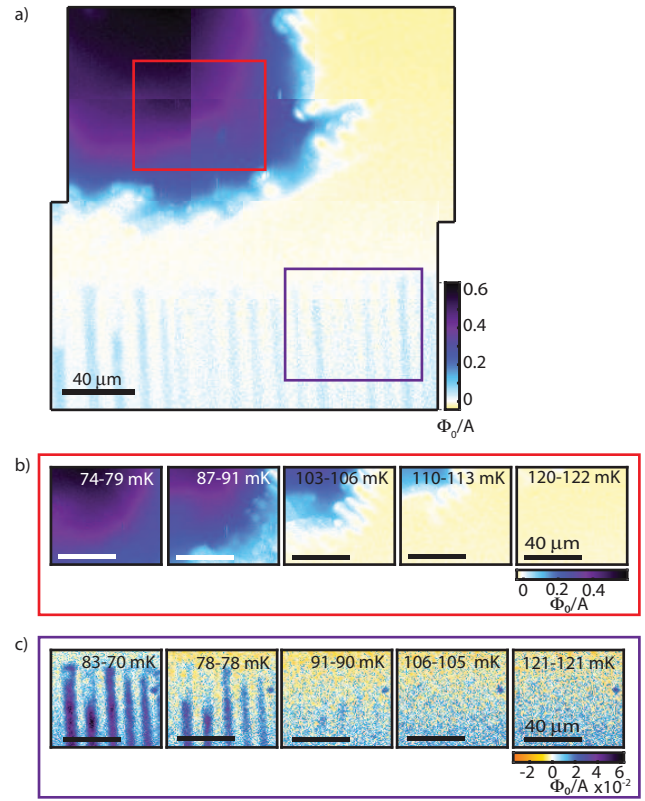}
\caption{\label{fig:Fig1} Sample A exhibits spatially inhomogeneous diamagnetism even at the lowest measurement temperatures. a) composite image of scans taken at 63 mK to 71 mK.  b) the region outlined with a red box in a) imaged repeatedly as a function of temperature; c) the region outlined with a purple box in a) imaged repeatedly as a function of temperature. The temperatures noted on the images in b) and c) were measured at the sample thermometer just before and just after scanning. Note: in these images, diamagnetic (i.e. superconducting) susceptibility is plotted with a \emph{positive} sign. Blue and purple indicate diamagnetism.}
\end{figure}

Two of the three areas imaged in Sample A did not exhibit diamagnetism down to the lowest temperatures examined (82-73 mK and 71-73 mK, respectively), as determined by examining measurements of the susceptibility as a function of height and confirming that the scanned areas were featureless. The third area [Fig.~\ref{fig:Fig1}(a)], measuring approximately $214 \times 224$ ${\mu}m^{2}$, contained two regions of superconductivity separated by a region tens of ${\mu}$m wide that did not superconduct down to at least 71 mK. 

The two superconducting regions in the third area of Sample A exhibited strikingly different spatial motifs. One region, in the upper left of Fig.~\ref{fig:Fig1}(a), exhibited spatially continuous and comparatively strong diamagnetism at the lowest temperatures that evolved into diagonally oriented tendrils of diamagnetism at temperatures close to $T_{c}$ [Fig.~\ref{fig:Fig1}(b)]. The other region,  in the lower half of Fig.~\ref{fig:Fig1}(a), consisted of a comb-like array of thin, diamagnetic features that remained separated by normal material down to at least 71 mK [Fig.~\ref{fig:Fig1}(c)]. For both superconducting regions, the orientations of the thin regions of diamagnetism are similar to possible orientations of tetragonal twin boundaries in STO, suggesting that $T_{c}$ is locally enhanced by the tetragonal domain structure of the STO, as has been seen in a different STO-based two-dimensional superconductor\cite{NoadPRB16}.

\begin{figure}
\centering
\includegraphics{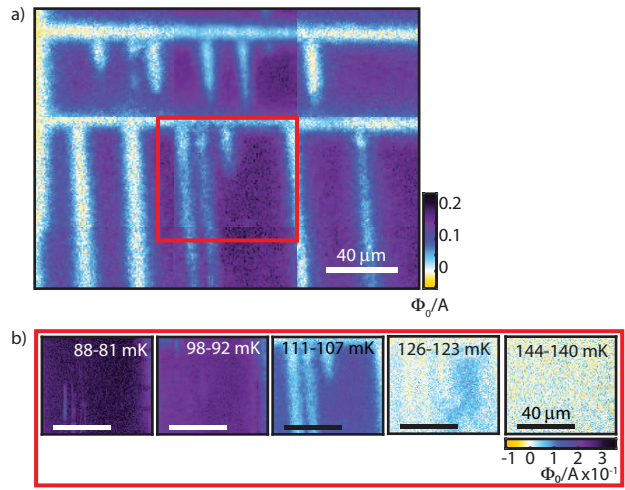}
\caption{\label{fig:Fig2}Sample B exhibits spatially inhomogeneous diamagnetism with a different pattern of variation from that observed in Sample A. a) composite image of scans taken at 102 mK to 116 mK on Sample B. b) the region outlined with a red box in a) imaged repeatedly as a function of temperature. The temperatures noted on the images in b) were measured at the sample thermometer just before and just after scanning. Note: in these images, diamagnetic (i.e. superconducting) susceptibility is plotted with a \emph{positive} sign. Blue and purple indicate diamagnetism.}
\end{figure}

In contrast to the large degree of spatial variation in Sample A, all three areas studied on Sample B showed similar patterns of diamagnetism. At intermediate temperatures, e.g.~102-117 mK for the area shown in Fig.~\ref{fig:Fig2}(a), narrow patches oriented at approximately $90^{\circ}$ to one another became normal while the majority of the scanned area remained superconducting to higher temperatures [Fig.~\ref{fig:Fig2}(b)]. The orientations of the edges of the regions with higher $T_{c}$ are suggestive of a subset of possible STO twin boundary orientations. Narrow features along former cubic $\langle$100$\rangle$ could also be due to dislocations in the STO. 

\section{\label{sec:discussion}Discussion}
The regions of locally enhanced $T_{c}$ that we observe in the optimally treated sample, Sample A, are generally thin (similar in width to the approximately 3 $\mu$m-diameter pickup loop of the SQUID) and are oriented in directions that are consistent with tetragonal domain boundaries in the STO. The features that are oriented at approximately $\pm45^{\circ}$ to the scan axes in Sample A [Fig.~\ref{fig:Fig1}(a), upper left, and (b)] must either be due to boundaries between domains with $c$ axis in-plane or else a pathological tiling of sub-resolution domains. The features that are approximately parallel to the scan axes (i.e. approximately parallel to the high-temperature cubic axes of the STO) could be either narrow domains of a certain orientation or the boundaries between them.

In the reduced sample, Sample B, the phenomenology is somewhat different; the narrow regions have \emph{lower} $T_{c}$, rather than higher, and they cross each other. These features could have the same origin as those in A, albeit with a very different spontaneously formed pattern.  Neither aspect ratio effects \cite{MullerSolidStateCommun70} nor strain from mounting seem likely to explain the difference in spatial motifs between the two samples, since the samples had similar dimensions and were mounted side-by-side using the same materials. Differences in carrier densities due to the post-growth annealing conditions (Table~\ref{table:conditions}) could potentially place Samples A and B on opposite sides of the dome in $T_{c}$ \cite{CavigliaNature08} such that perturbations due to twin structure would change $T_{c}$ in opposite ways.

How might structural domains in the STO tune $T_{c}$ in the LAO/STO? At least some of the features of locally higher $T_{c}$ in Sample A must be occuring at domain boundaries, due to their orientation of approximately 45$^{\circ}$ to $\langle$100$\rangle$. 
We hypothesize that polar domain walls in the STO \cite{SaljePRL13, MaPRL16, FrenkelNatMater17} are responsible for some, if not all, of the variation in $T_{c}$ in the optimally treated sample. Under this scenario, polar domain walls would enhance $T_{c}$ via charge accumulation, either by reducing scattering through enhanced screening or by enhancing the superfluid density. 

For Sample B, a different origin is possible: that its narrow regions of lower $T_{c}$ are unrelated to tetragonal domains in the STO and instead come from other defects that are oriented along the former cubic axes, such as dislocations, which are possibly charged \cite{ThielPRL09}. Oxygen vacancies migrate and preferentially dwell upon dislocations in STO during annealing at elevated temperatures, e.g. $700^{\circ}$C or higher, in a reducing atmosphere \cite{SzotPRL02}. Images of etched, oxygen-deficient STO in Ref.~\cite{SzotPRL02} show striking $\langle$100$\rangle$-oriented patterns of dislocations that cross one another at right angles. The length scale of spacing between the lines of etch pits in \cite{SzotPRL02} are at least an order of magnitude too small (1 ${\mu}m$ scale) to match the features seen in Sample B, however, which are spaced by of order 10 $\mu$m, though it is possible that different crystals would have different dislocation densities. 

The scenario of well-oriented regions of elevated oxygen deficiency producing the observed change in $T_{c}$ should not be relevant for Sample A as it was annealed in conditions designed to minimize oxygen deficiency. Oxygen deficiency should be ``frozen in" at room temperature and below \cite{MerkleAngewChem08}; thermal cycling just below and just above the 105 K structural phase transition of STO \cite{UnokiJPSJ67} could help to distinguish between dislocations and structural domains.

More generally, the differences between the amount and type of $T_{c}$ variation in the two samples may have to do with differences in the thickness and screening ability of the respective electron systems. In the oxygen-deficient sample, Sample B, the oxygen vacancy electron system may extend into the STO over a thickness of several microns \cite{SzotPRL02}; in contrast, the polar catastrophe-dominated electron system of the optimally treated sample, Sample A, is understood to be confined to within several unit cells at the STO side of the interface. A thicker electron system should be less sensitive overall to details of the electrostatic potential or crystal quality \emph{of the interface}, whereas the thinner 2D polar catastrophe electron system should be much more sensitive to variations in the quality of the interface and of variations in the electrostatic landscape at the interface, such as those caused by polar domain walls in the STO.  

Previous observations of spatially varying $T_{c}$ in $\delta$-doped STO \cite{NoadPRB16} seemed most consistent with effects related to domain orientation, rather than domain boundaries, though the authors did not exclude the possibility that domain boundaries could contribute to variation in $T_{c}$ as well. Optimally grown and treated LAO/STO may be more susceptible to domain boundary effects because the electrons originate from the polar discontinuity rather than from fixed dopant atoms in the plane of the 2D superconductivity. While the orientation dependence of the dielectric constant in STO \cite{SakudoPRL71} likely affects the tuning of the electrostatic potential at the interface, the difference in the polarizability along $a$ versus $c$ is small compared to the difference between the unpolarized domains and polar domain boundaries. Therefore, the latter likely has a stronger impact on the accumulation of electrons at the interface.

Finally, we note that the relatively low critical temperatures measured in these experiments are consistent with the understanding that the $T_{c}$ as measured by transport presents the $T_{c}$ of the percolation path with the largest critical temperature.

\section{\label{sec:conclusion}Conclusion}
We have shown that STO twin boundaries are an inherent source of spatial inhomogeneity in LAO/STO heterostructures that modulate the superconducting transition temperature in the two-dimensional superconductivity hosted at the interface with characteristic, allowed orientations. Twin boundaries that manifest as features parallel to the $\langle$100$\rangle$ direction within the superconducting plane could, in principle, be moved with a back gate \cite{HonigNatMater13}, raising the possiblility of making electric-field-tunable superconducting-normal state junctions in LAO/STO. Control over twin boundaries and other defects that arise near the LAO/STO interface will be necessary for realizing the full promise of this two-dimensional electron system.

\begin{acknowledgements}
We thank Christopher Watson for feedback on the manuscript and Hans Boschker for discussions. This work was supported by the Department of Energy, Office of Science, Basic Energy Sciences, Materials Sciences and Engineering Division, under Contract DE-AC02-76SF00515. H.N. acknowledges support from a Stanford Graduate Fellowship and from Natural Sciences and Engineering Research Council of Canada Post-Graduate Scholarships.
\end{acknowledgements}



\end{document}